\documentclass{article}
\usepackage{graphicx} 
\usepackage{amsmath}
\usepackage{parskip}
\usepackage{amssymb}
\newcommand{\E}{\mathbb{E}} 
\usepackage{physics}
\usepackage{mathtools}
\usepackage{xcolor}
\usepackage{float}
\usepackage{graphicx}
\usepackage{caption, geometry}
\usepackage{pdflscape}
\usepackage[
backend=biber,
style=alphabetic,
sorting=ynt
]{biblatex}
\usepackage{algorithm2e}
\usepackage{authblk}
\usepackage[]{caption}
\usepackage{hyperref}
\allowdisplaybreaks

\usepackage{tikz}

\title{Informed Trading model with Classical and Quantum Random Walks}
\title{Informed trading model with quantum walks}
\author[1]{H.W. Shawn Liew}
\author[1,2]{Patrick Rebentrost}
\affil[1]{Centre for Quantum Technologies, National University of Singapore}
\affil[2]{School of Computing, National University of Singapore}

\begin{document}
\maketitle

\section*{Abstract}

The seminal Kyle model offers a framework to understand the price impact of an informed trader in a market model for a single asset. We study variants in which the asset's future value is modeled by a classical random walk and by a continuous-time quantum walk, respectively. With these assumptions, we analytically approximate and numerically identify the market equilibrium between risk-neutral market maker and informed trader. In the classical equilibrium, the informed trader possesses a linear strategy, whereas the market maker's pricing rule resembles a logistic curve. The informed trader's profit is also identified to grow sub-linearly with time $t$. In the quantum equilibrium, the informed trader's strategy forms multiple clusters and the pricing rule exhibits multiple jumps correspondingly. Despite the increasing non-linearity of the equilibrium, the informed trader's profit grows linearly with time $t$.

\section{Introduction}

Future asset prices are inherently uncertain and  even hypothesized to be unpredictable according to the efficient market hypothesis \cite{fama1970efficient,follmer2016stochastic}. Empirically, fluctuating asset prices can exhibit drifts due to the emergence of new information over time \cite{bouchaud2018trades}.
From the perspective of the limit order book, complex price changes can be attributed to the order flow, i.e., the bids and asks and their corresponding volumes over time \cite{bouchaud2018trades}.
A fraction of bids or asks could be explained by informed trading, i.e., these are orders submitted by traders having some information of the asset's future value. Rational informed traders are expected to submit orders that would maximize their profits. There is a considerable literature on the impact of informed traders in a variety of market models \cite{kyle1985continuous,friedrich2020deepinvestingkylessingle, bouchaud2018trades}. 

A first well-studied insider trading model is the Kyle model from 1985 \cite{kyle1985continuous}. This single-auction model consists of three entities: an informed trader with knowledge of the future price of the asset, noise traders that trade without direction, and a risk-neutral market maker that makes zero profit. The traders submit orders to the market maker which are then executed at a price that is dependent on the volume of the orders. Under Gaussian assumptions, the Kyle model admits an equilibrium that is linear in the total order volume, with the linear coefficient known as the Kyle lambda. Since the introduction of the model, many extensions have been considered such as non-normality assumptions \cite{insdertradingwithoutnormality, insidertrading_nonlinearequilibria, bullish_bearish}, multi-variate \cite{garcia2020multivariate}, and risk-adverse entities \cite{subrahmanyam1991risk}. It is an interesting open direction to understand the effects of quantum dynamics in the context of the Kyle model. A particular dynamical process to consider would be quantum walks. 

Quantum walks are a generalization of classical walks. Various notions of quantum walks have been studied extensively in the context of quantum computing \cite{kendon_quantumwalk, childs_quantumwalk}. The seminal work of \cite{childs2003} demonstrates an exponential advantage in quantum walks over their classical counterparts on glued-tree graphs. Quantum walk techniques also underlie many algorithms in the query model of quantum computing \cite{miklostriangle, ambainisdatastructure}. 
The incorporation of ideas from physics into finance and economics theory has a long history \cite{bouchaud2018trades}. Several works have argued for combining concepts from quantum mechanics with finance and economics theory \cite{quantumbrownian, ORRELL2020122928, bao2023fundamentaltheoremquantumasset}. In particular, some works have incorporated quantum walks into financial models \cite{quantumwalk_assetprices, quantumwalkoptions, quantumwalk_DeBacker:2025wbz, quantumwalk_orrell2021quantum}. 

In this work, we  consider the Kyle model with a continuous-time quantum walk. We start from an analytical form of the quantum walk on a line and consider a single iteration between the insider and the market maker adapting their responses. While we obtain an analytical functional form of the approximate equilibrium, we resort to numerical methods to obtain the converged equilibrium. In the numerical equilibrium, we observe characteristics arising from the quantum walk such as oscillations in the order volume and the market maker's pricing rule. In particular, in the linear regime the Kyle lambda shows a quadratically larger slope than for the classical walk, due to the difference of ballistic and diffusive transport \cite{Rebentrost_2009,Plenio_2008}. We believe this work is a first to consider financial insider information in a quantum setting and that further discussions about other quantum scenarios related to insider information are interesting for future work.  

\section{Preliminaries}

\subsection{Kyle Model}

The Kyle model \cite{kyle1985continuous, friedrich2020deepinvestingkylessingle, bouchaud2018trades} offers a framework to model the price impact of an informed trader in the market for a single asset. In the single-auction setting, there are two time steps indicated by $t=0,1$. At time $t=0$, the price of the asset is $p_0$. Before ordering begins, the trader knows the price of the asset at $t=1$. We denote the future price by $z$. Thereafter, both the noise traders and the informed trader submit their orders to the market maker. The market maker then executes the orders at a price that is dependent on the size of the combined orders. Subsequently, it is $t=1$ and the price of the asset is updated to $z$.

\begin{figure}[H]
\centering
\fbox{\includegraphics[width=1\linewidth]{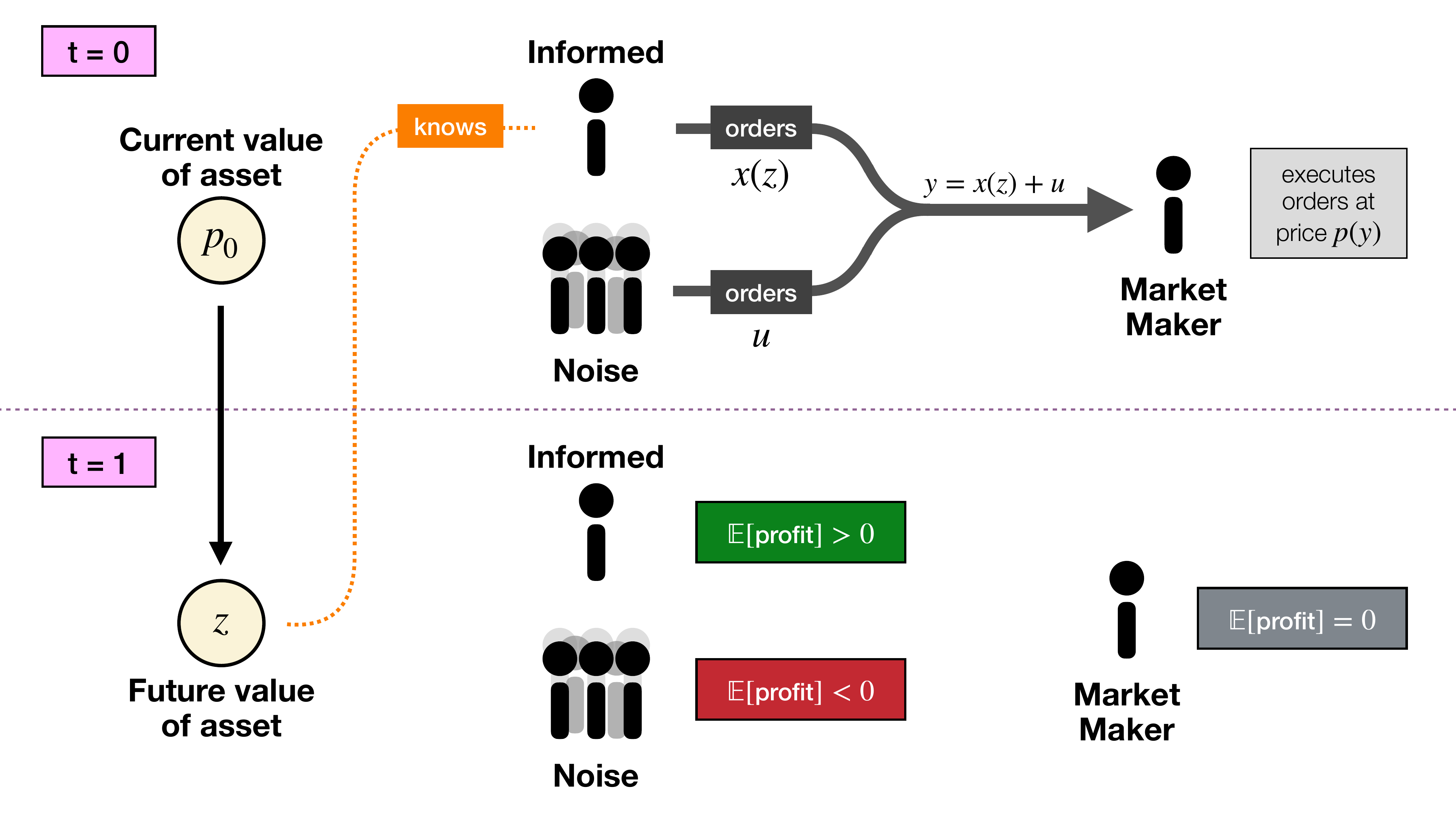}}
\caption{Illustration of the single-auction Kyle Model.}
\label{fig:kyle_model}
\end{figure}

We now formulate the objectives of each agent in this model. The noise traders' orders are driven by endogenous reasons and can be modeled by $u \sim N(0,\Sigma_u)$. Both the informed trader and the market maker have access to the description of $u$. 

The rational informed trader, wanting to maximize his advantage, selects an order that maximizes his profits. This can be formulated as
\begin{equation}
x(z) = \operatorname*{argmax}_x \E_u[\text{profit} | z ]  = \operatorname*{argmax}_x \E_u [x(z-p(x+u))].
\end{equation}

The market maker then receives the combined orders $y=x(z)+u$ and is unable to distinguish the informed trade from the noise trades. Since the competitive risk-neutral market is subjected to competition \cite{bouchaud2018trades}, the market maker thus makes zero profit on average. As such, the orders are executed at a competitive market-clearing price given by the pricing rule
\begin{align}
p(y) = \E_z[z|y] &= \int z f_{Z|Y}(z|y) dz = \frac{\int z f_{Y|Z}(y|z) f_z(z) dz}{f_Y(y)} = \frac{\int z f_u(y-x(z))f_z(z)dz}{\int f_u (y-x(z))f_z(z) dz}.
\label{eq:pricing_rule_def}
\end{align}

Having characterized the individual agent's optimization problem, we define the market equilibrium as the fixed-point of a pair of functions $(x^*, p^*)$ where $x^*$ corresponds to the informed trader's profit-maximizing strategy and $p^*$ corresponds to the market maker's zero-profit pricing rule.

\subsubsection{Kyle model with normality assumptions}

The noise traders' orders $u$ are described by a Gaussian random variable.
Suppose that the future price of the asset $z$ is also described by a Gaussian random variable $z \sim N(p_0, \Sigma_0)$. With these assumptions, the equilibrium pricing rule takes on a linear form specified by $p(y) = p_0 + \lambda y$ where $p_0$ is the initial price of the asset and $\lambda$ is the Kyle lambda which quantifies the price impact \cite{kyle1985continuous}. The informed trader's strategy then becomes
\begin{equation}
x(z) = \operatorname*{argmax}_x \E[x(z-p(x+u)] = \operatorname*{argmax}_x\E[x(z-p_0-\lambda x-\lambda u)] = \frac{z-p_0}{2\lambda}.
\end{equation}

Correspondingly, the market maker's clearing price is given by
\begin{equation}
p(y) = \E[z|y=x+u] = \E[z] + \frac{cov(z,y)}{var(y)}(y-\E[y]) = p_0 + \underbrace{\frac{\Sigma_0 / 2\lambda}{\Sigma_0/4\lambda^2\Sigma_u}}_{=\lambda}y.
\end{equation}

We can then solve $\lambda = \frac{\Sigma_0 / 2\lambda}{\Sigma_0/4\lambda^2\Sigma_u}$ to obtain $\lambda = \frac{1}{2}\sqrt{\frac{\Sigma_0}{\Sigma_u}}$. With this, the trader's average profit before knowing the future price $z$ (ex-ante expected profit) is given by 
\begin{equation}
\E_z \Big[ \E_u [\text{profit}|z] \Big] 
= \E_z \Big[x(z-p(x))\Big]
= \E_z \Big[ \frac{(z-p_0)^2}{4\lambda} \Big]
= \frac{\Sigma_0}{4\lambda}
= \frac{1}{2}\sqrt{\Sigma_0\Sigma_u}.
\end{equation}

As such, the unique market equilibrium is specified by $(x^*(z), p^*(y)) = \bigg( (z-p_0)\sqrt{\frac{\Sigma_u}{\Sigma_0}}, p_0 + \frac{1}{2} \sqrt{\frac{\Sigma_0}{\Sigma_u}}y \bigg)$.

\subsection{Classical Random Walk}

In our paper, we consider the classical random walk on a line with infinitely many nodes indexed by $j \in \mathbb{Z}$. The random walker's position $x$ at time $t=0$ is initialized at $j=0$. With each time step $t \in \mathbb{N}_0$, the classical walker randomly steps left or right with equal probability. The probability of finding the classical walker at node $j$ after $t$ time steps is thus given by

\begin{align}
\mathbb{P}(x=j|t) 
&= \frac{\mathbb{P}(x=j-1|t-1)  + \mathbb{P}(x=j+1|t-1)}{2} \nonumber \\
&= \begin{cases}
\binom{t}{(j+t)/2}(0.5)^t, & j \in \{-t,-t+2,\cdots,t-2,t\} \\
0, &\text{otherwise.}
\end{cases}
\end{align}

\begin{figure}[H]
\centering
\fbox{\includegraphics[width=\linewidth]{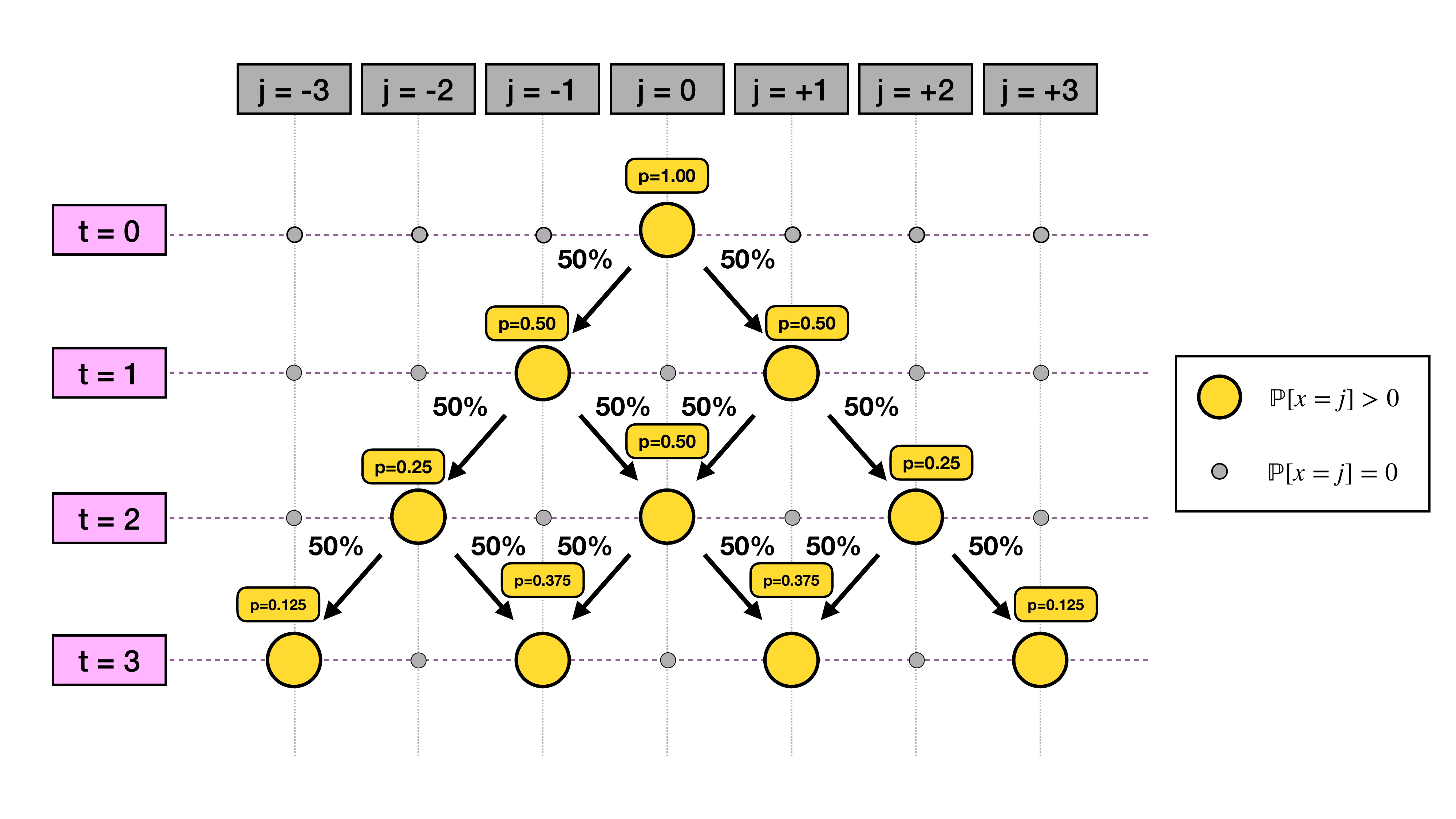}}
\caption{Illustration of classical random walk for $t=\{0,\cdots,3\}$ for $j=\{-3,\cdots,+3\}$.}
\label{fig:classical_random_walk}
\end{figure}

At large time steps $t \gg 1$, we can apply the Central Limit Theorem to approximate the Binomial probability distribution as a Gaussian probability distribution.

\subsection{Continuous-time Quantum Random Walk}

Similarly, we consider the continuous-time quantum walk \cite{Edward1998} on a line with infinitely many nodes indexed by $j \in \mathbb{Z}$. We represent the position of the walker with a quantum state which we initialize as $\ket{\psi(t=0)} = \ket{0}$. The basis states of the quantum state correspond to the possible nodes in which the random walker can occupy. Under the continuous-time quantum walk framework, the quantum state is subjected to the time-evolution of a Hamiltonian $H$ which encodes the balanced random walk for time $t \in \mathbb{R}_0^+$. We can formulate this as
\begin{equation}
\ket{\psi(t)} = \exp(-iHt/\hbar)\ket{\psi(t=0)} \quad \text{where} \quad 
H = \frac{1}{2} \sum_{j} \ket{j+1}\bra{j} + \ket{j-1}\bra{j}.
\end{equation}

We are able to solve $\ket{\psi(t)}$ analytically by first introducing the momentum eigenbasis
\begin{align}
\ket{k} &= \frac{1}{\sqrt{2\pi}}\sum_{j=-\infty}^{+\infty} e^{-ijk} \ket{j} 
\text{ where } k \in [-\pi, \pi] \quad \text{and} \quad
\ket{j} = \frac{1}{\sqrt{2\pi}}\int_{k=-\pi}^{+\pi} e^{ijk} \ket{k} dk.
\end{align} 

Applying the Hamiltonian $H$ defined above, we obtain
\begin{align}
H\ket{k} 
&= \frac{1}{2\sqrt{2\pi}} \sum_{j=-\infty}^{+\infty} e^{-ijk}[\ket{j+1} + \ket{j-1}]
= \frac{1}{2\sqrt{2\pi}} \sum_{j=-\infty}^{+\infty} e^{-ijk}[e^{ik} + e^{-ik}] \ket{j} = cos(k) \ket{k}.
\end{align}

Returning to our problem, we set $\hbar = 1$ and we then have
\begin{equation}
\ket{\psi(t)} 
= e^{-iHt}\ket{0} 
= e^{-iHt} \bigg[ \frac{1}{\sqrt{2\pi}} \int_{k=-\pi}^{+\pi} \ket{k} dk \bigg]
= \frac{1}{\sqrt{2\pi}} \int_{k=-\pi}^{+\pi} e^{-itcos(k)}  \ket{k} dk.
\end{equation}

To obtain the $j$-th coefficient of $\ket{\psi(t)}$, we apply the Jacobi-Anger expansion \cite{cuyt2008handbook} and the definition of the Dirac delta function to obtain
\begin{align}
c_j(t) 
= \langle j | \psi(t) \rangle 
&= \frac{1}{\sqrt{2\pi}} \int_{k=-\pi}^{+\pi} e^{-itcos(k)}  \langle j | k \rangle dk \nonumber \\
&= \frac{1}{2\pi} \int_{k=-\pi}^{+\pi} \Bigg[ \sum_{m=-\infty}^{+\infty} (-i)^m J_m(t) e^{imk} \Bigg]  e^{-ijk} dk \nonumber \\
&= \sum_{m=-\infty}^{+\infty} (-i)^mJ_m(t) \Bigg[ \frac{1}{2\pi} \int_{k=-\pi}^{\pi} e^{i(m-j)k} dk \Bigg] \nonumber \\
&= \sum_{m=-\infty}^{+\infty} (-i)^mJ_m(t) \delta_{m,j} 
= (-i)^j J_j(t).
\end{align}

where $J_j(t)$ is the Bessel function of the first kind.
Correspondingly, the probability of finding the quantum walker at node $j$ at time $t$ is given by
\begin{equation}
\mathbb{P}[x=j|t] = |(-i)J_j(t)|^2 = J_j^2(t).
\end{equation}

\begin{figure}[H]
\centering
\fbox{\includegraphics[width=.75\linewidth]{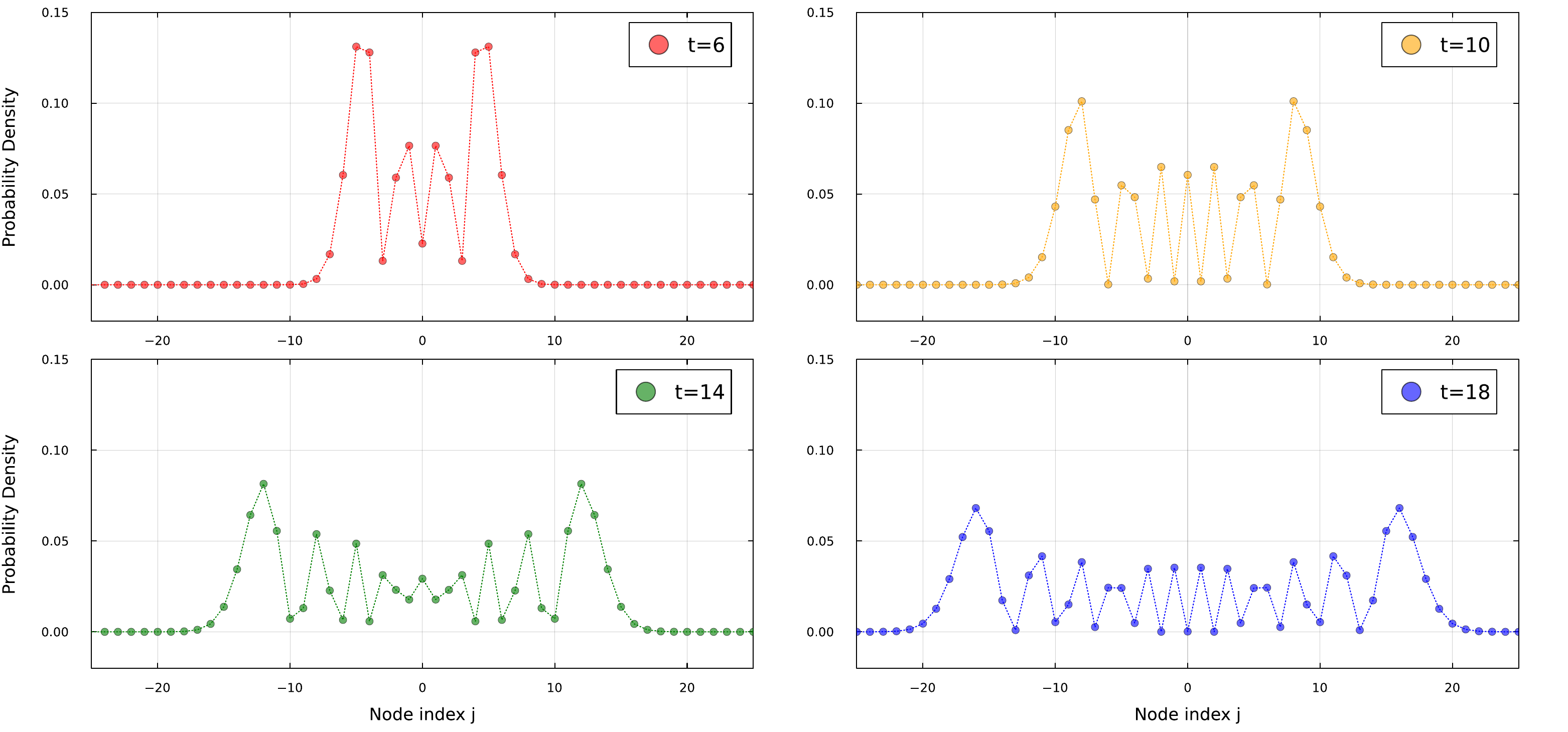}}
\caption{Probability Density Function of finding the quantum walker at node $j$ at time $t \in \{6,10,14,18\}$.}
\label{fig:bessel_pdf}
\end{figure}

\subsubsection{Relationship between time steps}

Similar to the classical random walk, we identify the relationship between the probabilities at time $t$ and at time $t-\epsilon$ where $\epsilon > 0$ and $\epsilon \ll 1$. In this case, we can use the Neumann's addition theorem \cite[\href{https://dlmf.nist.gov/10.23}{(10.23.2)}]{DLMF} to get
\begin{equation}
\mathbb{P}(x=j|t) = J_j^2(t) = [J_j(t-\epsilon + \epsilon)]^2 = \bigg[\sum_{k=-\infty}^{+\infty}J_{j-k}(t-\epsilon) J_k (\epsilon) \bigg]^2.
\end{equation}

To simplify the expression above, we use the following definition \cite{cuyt2008handbook}
\begin{equation}
J_m(x) = \sum_{k=0}^{+\infty} \frac{(-1)^k}{k!(m+k)!} \bigg(\frac{x}{2}\bigg)^{2k+m}.
\end{equation}

Since we have $\epsilon \ll 1$, it would be sufficient to consider just the first few values of $k$ to obtain an approximation of $J_m(x)$. 
\begin{align}
J_0(\epsilon) \approx  1 - \frac{\epsilon^2}{4}, \quad J_1(\epsilon) \approx \frac{\epsilon}{2}, \quad J_2(\epsilon) \approx \frac{\epsilon^2}{8} \quad \quad J_{-m}(\epsilon) &= (-1)^m J_m(\epsilon).
\end{align}

Putting all the terms together, we have
\begin{align}
J_n(t) = J_n(t - \epsilon + \epsilon) 
&= \sum_{k=-\infty}^{+\infty}J_{n-k}(t-\epsilon)J_k(\epsilon) \notag \\ 
&\approx J_n(t-\epsilon) + \bigg[ \frac{\epsilon}{2} \bigg] \bigg[  J_{n-1}(t-\epsilon) - J_{n+1}(t-\epsilon)  \bigg] \notag \\
&+ \bigg[ \frac{\epsilon^2}{8} \bigg] \bigg[ J_{n-2}(t-\epsilon) + J_{n+2}(t-\epsilon) - 2J_n(t-\epsilon)\bigg]
\end{align}

We can substitute the above to compute the corresponding probability
\begin{align}
\mathbb{P}(x=j|t) 
= J_j^2(t) 
&\approx J_j^2(t-\epsilon) + \epsilon J_j(t-\epsilon) \big[ J_{j-1}(t-\epsilon) - J_{j+1}(t-\epsilon) \big] \nonumber \\
&+ \frac{\epsilon^2}{4} \big[ J_{j-1}^2(t-\epsilon) + J_{j+1}^2(t-\epsilon) - 2J_{j-1}(t-\epsilon)J_{j+1}(t-\epsilon)\big] \\
&+ \frac{\epsilon^2}{4} J_j(t-\epsilon) \big[ J_{j-2}(t-\epsilon) + J_{j+2}(t-\epsilon) - 2J_j(t-\epsilon)\big]
\end{align}

As such, we have
\begin{align}
\mathbb{P}(x=j|t) 
&\approx \mathbb{P}(x=j|t-\epsilon) + \frac{\epsilon^2}{4}[\mathbb{P}(x=j-1|t-\epsilon) + \mathbb{P}(x=j+1|t-\epsilon)] \nonumber \\
&+ \epsilon J_j(t-\epsilon) [ J_{j-1}(t-\epsilon) - J_{j+1}(t-\epsilon)] - \frac{\epsilon^2}{2}J_{j-1}(t-\epsilon) J_{j+1}(t-\epsilon) \notag \\
&+ \frac{\epsilon^2}{4} J_j(t-\epsilon) \big[ J_{j-2}(t-\epsilon) + J_{j+2}(t-\epsilon) - 2J_j(t-\epsilon)\big] \label{eq:quantum_walk_final}
\end{align}

In Equation (\ref{eq:quantum_walk_final}), the first line resembles the expression obtained in the classical random walk setting. The second and third line itself provides two takeaways. The time evolution of the random walker deals with the underlying probability amplitude $J_n(t)$ in the quantum setting rather than $\mathbb{P}(x=j|t)$ as in the classical setting. Additionally, the quantum setting includes not only the interaction between the current node and its neighbors but also the interactions among all its neighbors.

\section{Kyle Model with Classical and Quantum Walks}

Our paper considers a variant of the Kyle model in which $z$, the future price of the asset, is modeled by either a classical random walk or a continuous-time quantum walk. We first initialize the random walker at the center of a line with infinitely many nodes indexed by $j \in \mathbb{Z}$. Each node is then associated with a possible future price of the asset $z_j$. In our case, we set $z_j = p_0 + sj$ where $s$ is a constant specifying the step size. As such, the probability that the future price of the asset will become $z_j$ is equal to the probability of finding the walker at the node $j$. In the setting of the classical random walk, we have
\begin{equation}
f_z(z_j) = \mathbb{P}(z = z_j|t) = \mathbb{P}(x=j|t) = 
\begin{cases}
\binom{t}{(j+t)/2}(0.5)^t, & j \in \{-t,-t+2,\cdots,t-2,t\} \\
0, &\text{otherwise.}
\end{cases}
\end{equation}

For the continuous-time quantum random walk, we have
\begin{equation}
f_z(z_j) = \mathbb{P}(z = z_j|t) = \mathbb{P}(x=j|t) = J_j^2(t).
\end{equation}

We seek to identify the market equilibrium $(x^*, p^*)$ for both settings.

\subsection{Mathematical formulation}

As the future price of the asset can now only take discrete values $z_j$, we reformulate the objectives of both the informed trader and the noise traders. The informed trader has perfect knowledge of $z_j$ and intends to maximize his profits by submitting the order $x(z)$. This is given by
\begin{align}
x(z) 
&= \operatorname*{argmax}_{x} \E[\text{profit} | z] 
= \operatorname*{argmax}_{x} \E[x(z-p(y))] \nonumber \\
&= \operatorname*{argmax}_{x} \Big[ xz - x \int_{u=-\infty}^{+\infty} p(x+u) f_u(u) du \Big] \nonumber \\
&\xrightarrow{\text{discretize}} \operatorname*{argmax}_{x}
\Bigg[ xz - x \frac{\sum_j p(x+sj) \exp(-(sj)^2 / 2\Sigma_u)}{\sum_j \exp(-(sj)^2/2\Sigma_u)} \Bigg].
\label{eq:discretized_informed_trader}
\end{align}

Correspondingly, the competitive risk-neutral market maker makes zero profit and would have to execute the orders according to the pricing rule $p(y)$ given by 
\begin{align}
p(y) 
&= \E[z|y] 
= \frac{\int z f_u(y-x(z)) f_z(z) dz}{\int f_u(y-x(z)) f_z(z) dz} \nonumber \\
&\xrightarrow{\text{discretize}} \frac{ s \sum_j z_j f_u(y-x(z_j))f_z(z_j)}{ \displaystyle s\sum_j f_u(y-x(z_j))f_z(z_j)} \nonumber \\
&= p_0 + s \frac{ \displaystyle \sum_j j \, \exp\Bigg(\frac{-(y-x(p_0+sj))^2}{2\Sigma_u}\Bigg) f_z(z_j)}{\displaystyle \sum_j \exp\Bigg(\frac{-(y-x(p_0+sj))^2}{2\Sigma_u} \Bigg) f_z(z_j)},
\label{eq:discretized_pricing_rule}
\end{align}

where we made substitutions (i) $z_j \leftarrow p_0 + sj$ and (ii) $f_u(y-x(z_j)) \leftarrow \frac{1}{\sqrt{2\pi\Sigma_u}} \exp\Big( \frac{-(y-x(p_0+sj))^2}{2\Sigma_u} \Big)$. We then have
\begin{equation*}
f_z(z_j)_{\text{classical}} =
\begin{cases}
\binom{t}{(j+t)/2}(0.5)^t, & j \in \{-t,-t+2,\cdots,t-2,t\} \\
0, &\text{otherwise}
\end{cases} 
\quad \quad \quad
f_z(z_j)_{\text{quantum}} = J_j^2(t).
\end{equation*}

\subsection{Analytical iteration for quantum walk}

In the continuous-time quantum walk setting, we first attempt an analytical approach to the problem. We assume that the market maker has a linear pricing rule given by $p(y) = p_0 + \lambda y$. Correspondingly, the informed trader submits the order $x(z) = (z-p_0)/2\lambda$. Upon doing so, we can compute the market maker's updated pricing rule.

\begin{align}
p(y) &\rightarrow
\E[p_0+sj \,|\, y] = p_0 + s \E[j|y] \notag \\
&= p_0 + s \frac{ \displaystyle \sum_j j \, \exp\Big(\frac{-(y-sj/2\lambda)^2}{2\Sigma_u}\Big) J_j^2(t)}{ \displaystyle \sum_j \exp\Big(\frac{-(y-sj/2\lambda)^2}{2\Sigma_u} \Big) J_j^2(t)} \notag \\
&= p_0 + s\, 
\frac{ \displaystyle \sum_{j=-\infty}^{-1} j J_{j}^2(t) \exp\bigg(\frac{-(y-sj/2\lambda)^2}{2\Sigma_u} \bigg) + \sum_{j=1}^{+\infty} j J_{j}^2(t) \exp\bigg(\frac{-(y-sj/2\lambda)^2}{2\Sigma_u} \bigg)}
{\displaystyle \sum_{j=-\infty}^{-1} J_{j}^2(t) \exp\bigg(\frac{-(y-sj/2\lambda)^2}{2\Sigma_u} \bigg) + J_0^2(t) \exp\bigg(\frac{-y^2}{2\Sigma_u}\bigg) + \sum_{j=1}^{+\infty} J_{j}^2(t) \exp\bigg(\frac{-(y-sj/2\lambda)^2}{2\Sigma_u} \bigg)} \notag \\
&= p_0 + s\,
\frac{\displaystyle \sum_{j=1}^{+\infty} j J_{j}^2(t) \exp\bigg( \frac{-s^2j^2}{8\lambda^2\Sigma_u} \bigg) \bigg[ \exp\bigg( \frac{sjy}{2\lambda \Sigma_u} \bigg) - \exp\bigg(- \frac{sjy}{2\lambda \Sigma_u} \bigg) \bigg]}
{\displaystyle J_0^2(t) + \sum_{j=1}^{+\infty} J_{j}^2(t) \exp\bigg( \frac{-s^2j^2}{8\lambda^2\Sigma_u} \bigg) \bigg[ \exp\bigg( \frac{sjy}{2\lambda \Sigma_u} \bigg) + \exp\bigg(- \frac{sjy}{2\lambda \Sigma_u} \bigg) \bigg]} \hspace{5mm} \notag \\ &(\text{where we used } J_j^2(t) = J_{-j}^2(t)) \notag \\
&= p_0 + 2s\,
\frac{\displaystyle \sum_{j=1}^{+\infty} j J_{j}^2(t) \exp\bigg( \frac{-s^2j^2}{8\lambda^2\Sigma_u} \bigg) \sinh\bigg( \frac{sjy}{2\lambda \Sigma_u} \bigg)}
{\displaystyle J_0^2(t) + 2\sum_{j=1}^{+\infty} J_{j}^2(t) \exp\bigg( \frac{-s^2j^2}{8\lambda^2\Sigma_u} \bigg) \cosh\bigg( \frac{sjy}{2\lambda \Sigma_u} \bigg)}.
\label{eq:analytical_pricing_rule}
\end{align}

\begin{figure}[H]
\centering
\fbox{\includegraphics[width=.9\linewidth]{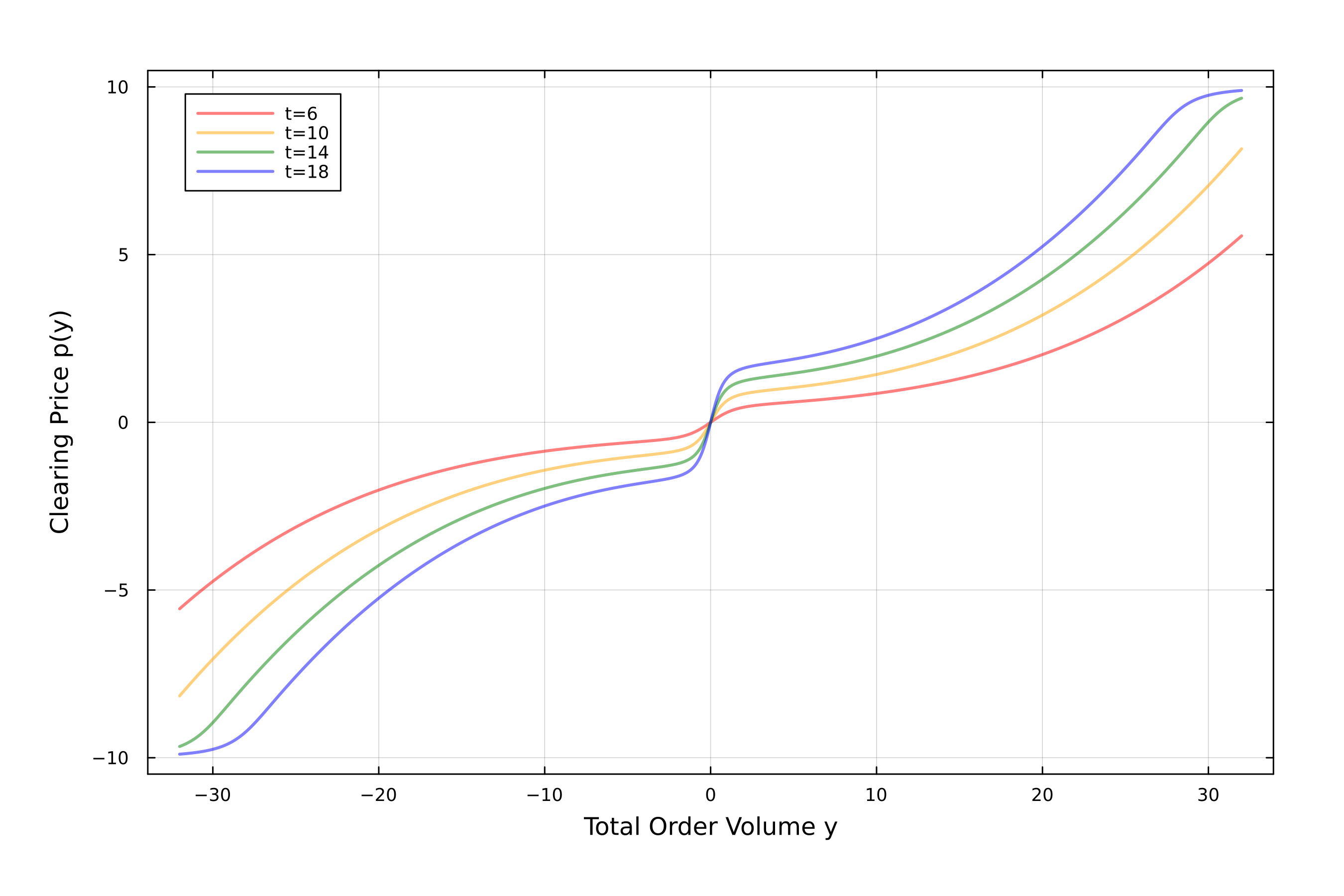}}
\caption{Plot of the Clearing Price in Equation (\ref{eq:analytical_pricing_rule}) against $y$ for $t \in \{6,10,14,18\}$.}
\label{fig:analytical_clearing_price}
\end{figure}

From Equation (\ref{eq:analytical_pricing_rule}) and illustrated by Figure \ref{fig:analytical_clearing_price}, it is clear that the market equilibrium is not linear. We can attempt to understand how the price impact varies with $y$ in this setting. To simplify our upcoming work, we define the following
\begin{equation*}
A_j := J_j^2(t) \exp \bigg(\frac{-s^2j^2}{8\lambda^2 \Sigma_u} \bigg), \quad B := \frac{s}{2\lambda\Sigma_u},
\end{equation*}
and
\begin{equation*}
N(y):= \sum_{j=1}^{+\infty} j A_j \sinh(Bjy), \quad D(y):= J_0^2(t) + 2\sum_{j=1}^{+\infty} A_j \cosh(Bjy). 
\end{equation*}

Correspondingly, we then have
\begin{align*}
N'(y) &= \sum_{j=1}^{+\infty} j^2 A_j B \cosh(Bjy) = BM(y) \quad \text{where } M(y):= \sum_{j=1}^{+\infty} j^2 A_j \cosh(Bjy) \\
D'(y) &= 2\sum_{j=1}^{+\infty} j A_j B \sinh(Bjy) = 2BN(y).
\end{align*}

The price impact of the updated pricing rule can now be written as
\begin{align}
p(y) &= p_0 + 2s\frac{N(y)}{D(y)} \nonumber \\
\quad \Rightarrow \, 
\frac{dp(y)}{dy} &= 2s \bigg( \frac{N'(y)D(y) - N(y)D'(y)}{D(y)^2} \bigg) = 2s \bigg( \frac{BM(y)D(y) - 2BN(y)^2}{D(y)^2} \bigg)\\
&= \frac{s^2}{\lambda \Sigma_u} \bigg( \frac{M(y)}{D(y)} - 2\frac{N(y)^2}{D(y)^2} \bigg).
\label{eq:pricing_rule_derivative}
\end{align}

Notice that if we were to obtain the expression for 
\begin{align}
\E[j^2|y]
&= \frac{ \displaystyle \sum_j j^2 \, \exp\Big(\frac{-(y-sj/2\lambda)^2}{2\Sigma_u}\Big) J_j^2(t)}{ \displaystyle \sum_j \exp\Big(\frac{-(y-sj/2\lambda)^2}{2\Sigma_u} \Big) J_j^2(t)} = 2 \frac{ \displaystyle \sum_{j=1}^{+\infty} j^2 A_j \cosh(Bjy) }{ J_0^2(t) + \displaystyle 2\sum_{j=1}^{+\infty} A_j \cosh (Bjy) } = 2 \frac{M}{D}.
\end{align}

We can now find an expression for $\text{Var}(j|y)$ to substitute into Equation (\ref{eq:pricing_rule_derivative})
\begin{align}
\text{Var}(j|y) = \E[j^2|y] - (\E[j|y])^2 = 2 \bigg(\frac{M(y)}{D(y)} - 2\frac{N(y)^2}{D(y)^2}\bigg)
\quad \Rightarrow \quad 
p'(y) = \frac{dp(y)}{dy} = \frac{s^2}{2\lambda\Sigma_u} \text{Var}(j|y).
\label{eq:analytical_price_impact_var}
\end{align}

The final expression in Equation (\ref{eq:analytical_price_impact_var}) suggests that the price impact is largest when the market maker has low confidence in the actual value of the asset's future price and vice versa. From this sub-section, it is clear that this analytical approach becomes increasingly complicated and will become intractable. As such, we resort to numerical methods.

\section{Numerical results for Kyle Model with Classical Random Walk}

We now numerically identify the market equilibrium $(x^*,p^*)$ in the setting where the future price of the asset is modeled by a classical random walk parameterized by the time step $t \in \mathbb{N}_0$. In our computation, we considered $t \in \{1,\cdots,23\}$. This range was selected to enable a comparison of the results in this section with the results of the Kyle model in the continuous-time quantum setting. In the quantum regime, the algorithm had failed to identify an equilibrium for large values of $t$. As such, we presented results for only $t \in \{1,\cdots,23\}$ in the quantum setting; and imposed the same range of $t$ for this section.

\subsection{Numerical Methods}

To obtain the market equilibrium, we employed a fixed-point iterator described in Algorithm \ref{alg:clearing_price_equilibrium}. For this algorithm, we specify $u \sim N(0, \Sigma_u = 1)$ and step size $s=0.1$ as in $z_j = p_0 + sj$. We then specify the set of hyperparameters which are required for the algorithm. These hyperparameters are as described. 

\begin{enumerate}
\item {Discretization.} The aggregate order flow $y$ is discretized over the range starting from $-30+y_{\text{origin}}$ with step size $y_{\text{step}}$ up to an upper bound of $+30$. The resulting grid is defined as:
\begin{equation*}
\mathcal{Y} = \{ y_k = -30 + y_{\text{origin}} + k \cdot y_{\text{step}} \, | \, k = 0, 1, ..., N \},
\end{equation*}
where the number of steps is $N = \left\lfloor \frac{60 - y_{\text{origin}}}{y_{\text{step}}} \right\rfloor$ and $\max(\mathcal{Y}) \le 30$. We set $y_{\text{origin}} = 0$ and $y_{\text{step}} = 0.01$. 

\item {Initial Pricing Rule.} The initial pricing rule is defined as 
\begin{equation*}
\widetilde{p}(y)_0 = \displaystyle \sum_{i=1}^3 \alpha_i \Big[ tanh(\beta_i y - \gamma_i) + tanh(\beta_i y + \gamma_i) \Big],
\end{equation*}
where $\alpha_1 = 0.08, \beta_1 = t$, and $\gamma_1 = 0$. The other hyperparameters $\alpha_i, \beta_i, \gamma_i$ for $i \in \{2,3\}$ were set to $0$.

\item {Anderson Acceleration with dampening.} 
The Anderson acceleration method \cite{anderson_acceleration} accelerates the fixed-point iterator employed in Algorithm \ref{alg:clearing_price_equilibrium}. At each iteration, it retrieves the history of the $m$ most recent residual vectors $\{g_i\}_{i=0}^m$ and constructs the residual matrix $G := [g_1 \,\, g_2 \,\, \cdots \,\, g_m]$. The weights $\{\alpha_i\}_{i=0}^m$ are computed such that $||G \alpha||_2$ is minimized. These weights are combined with the damping parameter to compute the element for the next iteration. In our setting, we have
\begin{align*}
\tilde{p}_{k+1} 
&= \text{AndersonStep}(m, \beta, \{p_k\}, \{\tilde{p_k}\}, \{r_k\}) \\
&= \beta \sum_{i=0}^{m_k} \alpha_i \, p_{k-m_k+i} +
(1-\beta) \sum_{i=0}^{m_k} \alpha_i \, \tilde{p}_{k-m_k+i},
\end{align*}
where $m_k = \min(m,k)$. We set $m = 10$ and $\beta = 1.0$.

\item {Asset future prices.} Recall that we set $z_j = p_0 + sj$. In the classical random walk setting, we have $\mathbb{P}[z=z_j] > 0$ for $j \in \{-t,-t+2,\cdots,t-2, t\}$. As such, we only consider nodes with non-zero probability which we denote by 
\begin{equation*}
\{z_j\}_{-M\,:\, \Delta j \,:\, M} := \{ -M, -M+\Delta j, -M+2\Delta j, \cdots, M-\Delta j, M\},
\end{equation*}
where we set $M = t$ and $\Delta j = 2$.

\end{enumerate}

\vspace{3mm}

\RestyleAlgo{ruled} 
\begin{algorithm}[H]
\caption{Market Equilibrium fixed-point iterator using optimal hyperparameters}
\label{alg:clearing_price_equilibrium}

\SetKwInOut{Input}{Input}
\SetKwInOut{Output}{Output}
\Input{Residue tolerance $\epsilon = 10^{-4}$, Number of iterations $N=1000$, Parameters $\boldsymbol{\theta}$}
\Output{Informed Trader's Strategy $x^*$, Market Maker's Pricing Rule $p^*$, Residue $r_{\max}$}

\BlankLine
\tcp{Initialize using the parameters}
Initialize spatial grid $\mathbf{y}$ and initial pricing rule $\widetilde{p}_{k=1}$ \\
Initialize history buffers $P = [], \tilde{P} = [], R = []$ \\

\BlankLine
\tcp{Fixed-Point Iterator}
\For{$\text{k} \leftarrow 1$ \KwTo $N$}{
Compute informed trader's order flow $x_k$ for $\{z_j\}_{-M\,:\, \Delta j \,:\,M}$ from $\widetilde{p}_k$ \\
Compute market maker's pricing rule $p_k$ from $x_k$ \\
Compute residual vector $r_k \leftarrow p_k - \tilde{p}_k$ \\
Update history buffers with $(p_k, \tilde{p}_k, r_k)$ \\
Compute $\widetilde{p}_{k+1} \leftarrow \text{AndersonStep}(m, \beta, P, \tilde{P}, R)$ \\
}
Compute final residue: $r_{\max} \leftarrow \operatorname*{max}_{y} | \widetilde{p}_N - p_N|$ \\

\BlankLine
\BlankLine
\Return $p_N^*, x_N^*, r_{\max}$
\label{alg:clearing_price_equilibrium}
\end{algorithm}  
\vspace{3mm}

\begin{figure}[H]
\centering
\fbox{\includegraphics[width=0.92\linewidth]{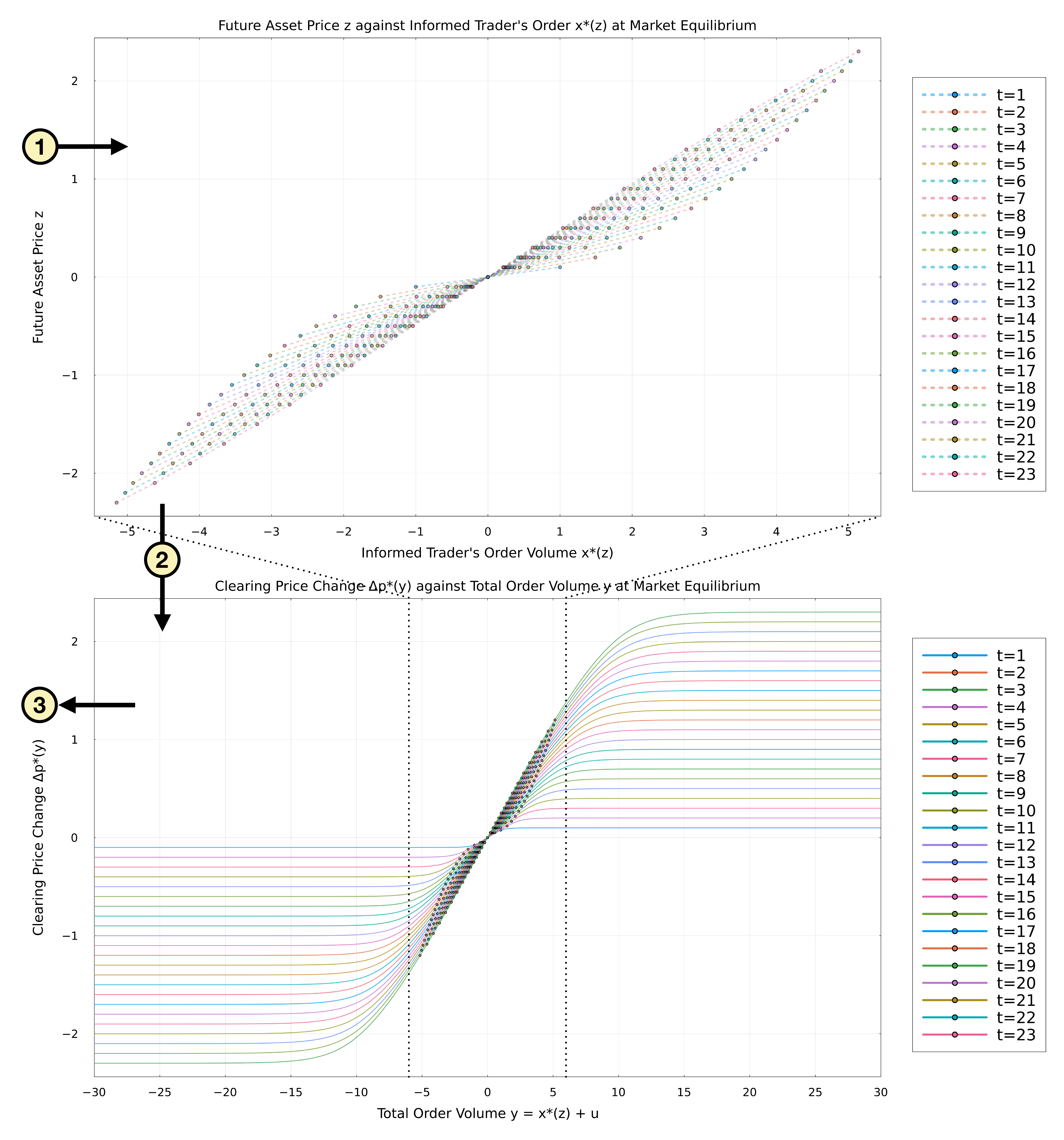}}
\caption[]{
Numerically-obtained Market Equilibrium $(x^*,p^*)$ curves for the Kyle model in which the future price of the asset is modeled by a classical random walk. We computed the equilibrium for $t \in \{1,\cdots,23\}$. The yellow circles indicate the sequence of events in the Kyle model. Initially, the value of $t$ is fixed. 
\begin{tikzpicture}
\node[draw, circle, inner sep=1pt,minimum size=1.0em] at (0,0) {1};
\end{tikzpicture}
The informed trader is then made known of the future price of the asset $z$. 
\begin{tikzpicture}
\node[draw, circle, inner sep=1pt, minimum size=1.0em] at (0,0) {2};
\end{tikzpicture} Subsequently, the informed trader submits $x^*(z)$. \begin{tikzpicture}
\node[draw, circle, inner sep=1pt, minimum size=1.0em] at (0,0) {3};
\end{tikzpicture} The market maker finally decides the clearing price based on the total order volume $y=x^*(z)+u$.
(Top) The plot represents the informed trader's equilibrium strategy $x^*(z)$ given a future asset value $z$. The dotted lines are only for visual aid. (Bottom) The plot represents the change in the market maker's pricing rule $p^*(y)$ for every total order volume $y$. We have defined $\Delta p^*(y)$ as $ p^*(y) = p_0 + \Delta p^*(y)$. The circles on the plot indicate the possible orders of the informed trader.
}
\label{fig:classical_equilibrium_combined}
\end{figure}

With the hyperparameters specified above, Algorithm \ref{alg:clearing_price_equilibrium} converged to a market equilibrium $(x^*,p^*)$ for every value of $t \in \{1,2,...,23\}$. The results are presented in Figure \ref{fig:classical_equilibrium_combined}.

\newpage
\begin{figure}[H]
\centering
\fbox{\includegraphics[width=0.9\linewidth]{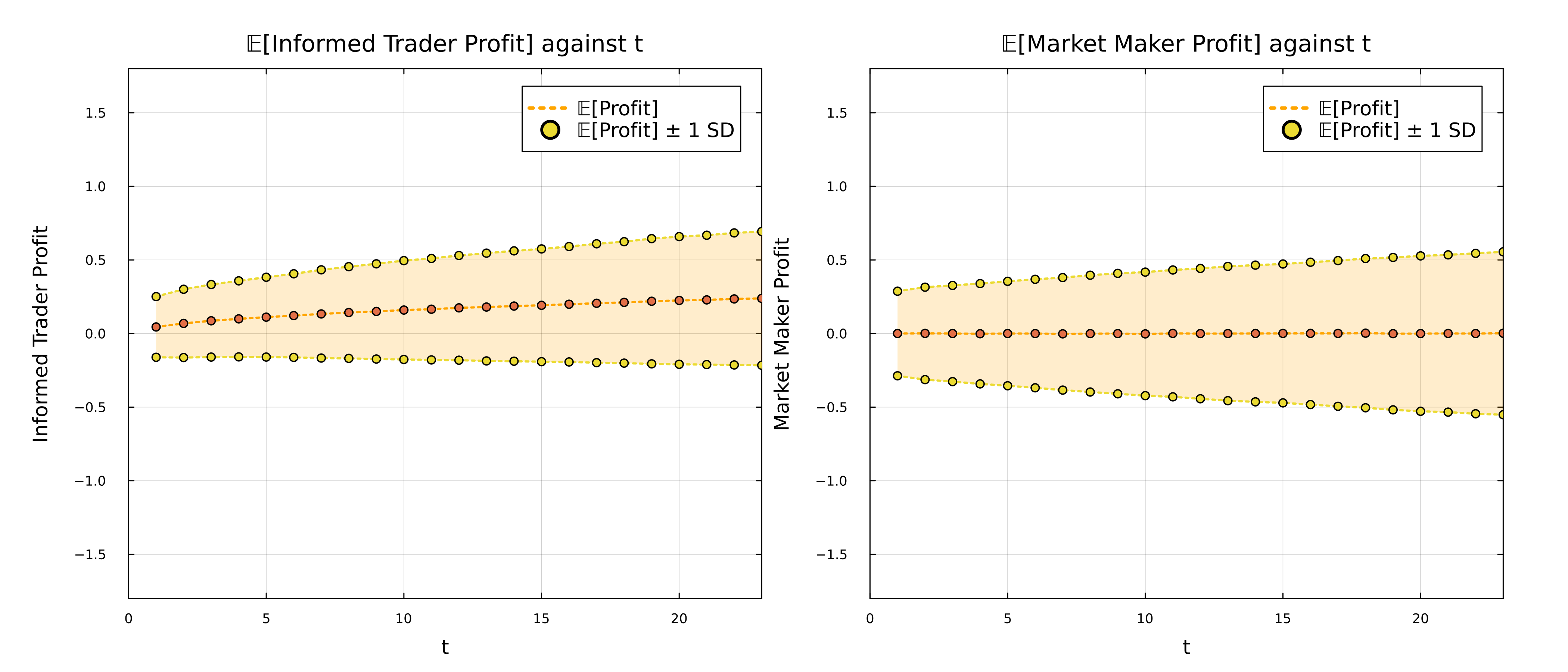}}
\caption{Average profit per round for $t \in \{1,...,23\}$ based on $200,000$ rounds of the Kyle model for (Left) Informed Trader (Right) Market Maker.}
\label{fig:classical_combined_profit}
\end{figure}

\begin{figure}[H]
\centering
\fbox{\includegraphics[width=0.9\linewidth]{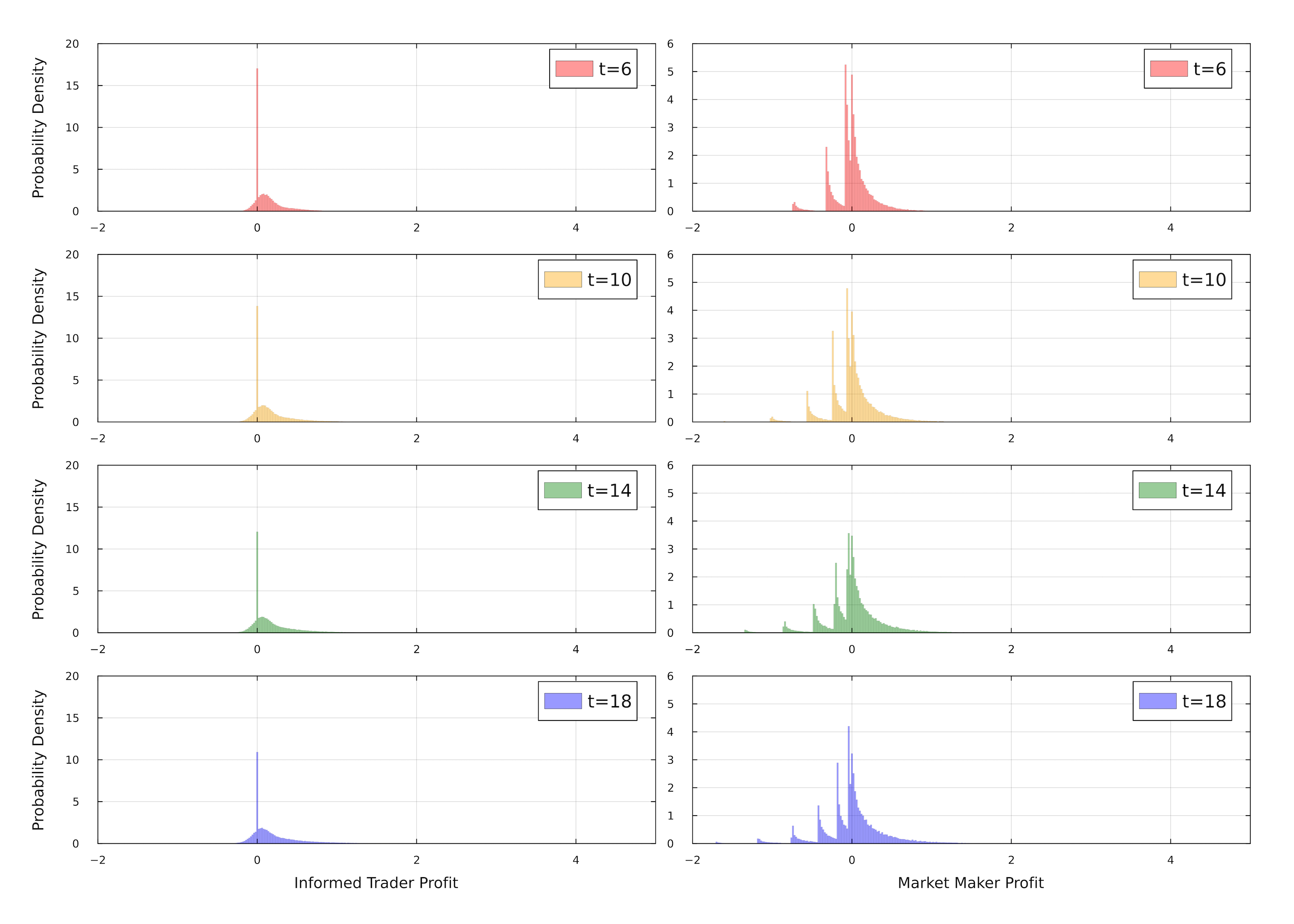}}
\caption{Probability density based on $200,000$ rounds of the Kyle model for $t \in \{6,10,14,18\}$ for \linebreak(Left Column) Informed Trader (Right) Market Maker.}
\label{fig:classical_profit_grid}
\end{figure}

\subsection{Discussions}

In the classical random walk setting, we found that the market equilibrium $(x^*, p^*)$ admits a linear informed trader strategy for $t \in \{1,\cdots 23\}$. We verified its linearity by fitting a linear regression model and found that $R^2 > 0.996$ for all the plotted values of $t$. The corresponding clearing price was also found to be linear at around $y=0$ and tapers off for large values of $|y|$ at $\Delta p^*(y) = st$: the minimum and maximum value of $z$ achievable in this model. From these curves, we also identified that the Kyle lambda for $t \in \{2,\cdots 23\}$ around $y=0$ grows sub-linearly with $t$ and can be described by $\lambda (t) \approx0 .0514 t^{0.491} + 0.00167$. The Kyle lambda of $t=1$ was an outlier and had a value of $\lambda(t=1) \approx 0.0100$. 

From Figure \ref{fig:classical_combined_profit}, the average profit per round for the informed trader was also identified to grow sub-linearly with $t$ at $\E[\text{informed trader profit}] \propto t^{0.69}$.

In Figure \ref{fig:classical_profit_grid}, we observe two interesting features. In the left column, a distinct spike occurs where the informed trader makes zero profit. This spike can be attributed to the case where $z=p_0$ in which the informed trader makes no trade and thus zero profit. Since $z=p_0$ is the most probable price of $z$ in the classical random walk setting, the magnitude of the spike is also correspondingly large. 

In the right column, the distribution of the market maker profits consists of multiple asymmetrical waveforms. Each asymmetrical waveform can be attributed to profits made at different values of $z$. The larger the value of $|z-p_0|$, the more loss the market maker incurs due to the informed trader, and hence the more left-shifted the waveform appears. The asymmetrical profile can then be attributed to the noise trades. Consider the case where $z=p_0$. The informed trader does not trade but the noise traders may submit a non-zero order volume. In this case, the market maker will execute the orders at a price $p(y)$ away from $z = p_0$. As such, the market maker is guaranteed to make at least zero profit from the noise traders. The amount of profit made would then be dependent on the volume of the noise trades which is described by a Gaussian distribution.

\section{Numerical results on the Kyle Model with continuous-time Quantum Random Walk}

Similarly, we seek to compute the market equilibrium $(x^*, p^*)$ in the setting where the future price of the asset is modeled by a continuous-time quantum walk parameterized by the time step $t \in \mathbb{R}_{\geq0}$. We only considered $t \in \{1,...,23\}$ to ensure that the results obtained in the quantum setting can be easily compared to those in the classical setting.

\subsection{Numerical Methods}

To identify the market equilibrium, we employed the fixed-point iterator described in Algorithm \ref{alg:clearing_price_equilibrium}. However, we observed empirically that the quality of the convergence was highly sensitive to the input hyperparameters. As such, we first used Algorithm \ref{alg:equilibrium_hyperparam_tuning} to perform hyperparameter fine-tuning using Optuna \cite{optuna_2019}. Upon obtaining an optimal set of hyperparameters, we use these hyperparameters as input for Algorithm \ref{alg:clearing_price_equilibrium}. The hyperparameters that were tuned are listed below.

\begin{enumerate}
\item {Discretization.} $y_{\text{origin}}$ and $y_{\text{step}}$. 
\item {Initial Pricing Rule.} $\alpha_i, \beta_i, \gamma_i$ for $i \in \{1,2,3\}$.
\item {Anderson Acceleration with dampening.}  $m_k$ and $\beta$.
\item {Asset terminal value prices.} $M$. We set $\Delta j = 1$.
\end{enumerate}

In this setting, we used $u \sim N(0,\Sigma_u = 1)$ and $s=0.1$ in $z_j=p_0+sj$ as well.

\RestyleAlgo{ruled} 
\begin{algorithm}[H]
\caption{Market Equilibrium Fixed-Point Iterator Hyperparameter Tuning}
\label{alg:equilibrium_hyperparam_tuning}

\DontPrintSemicolon
\SetKwInOut{Input}{Input}
\SetKwInOut{Output}{Output}

\Input{Residue tolerance $\epsilon = 10^{-4}$, Number of iterations $N=1000$, Max Optuna trials $N_{\text{optuna}}=100$}
\Output{Optimal Hyperparameters $\boldsymbol{\theta}^*$}

\BlankLine
$\mathcal{T} \gets \emptyset$         \tcp{Storage for completed Optuna trials}

\BlankLine
\tcp{Optuna Hyperparameter Exploration}
\For{$\text{trial} \gets 1$ \KwTo $N_{\text{optuna}}$}{
$\boldsymbol{\theta} \gets \text{Suggest Hyperparameters}(\{y_{\text{origin}}, y_{\text{step}}\}, \{\alpha_i, \beta_i, \gamma_i\}_{i=1}^3, \{m, \beta\}, \{M\})$\;
Initialize spatial grid $\mathbf{y}$ and initial pricing rule $\widetilde{p}_1$\;

\BlankLine
\tcp{Fixed-Point Iterator with Anderson Acceleration}
\For{$k \gets 1$ \KwTo $N$}{
Compute informed trader's order flow $x_k$ for $\{z_j\}_{j=-M}^{+M}$ from $\widetilde{p}_k$\;
Compute market maker's pricing rule $p_k$ from $x_k$\;
$\widetilde{p}_{k+1} \gets \text{AndersonStep}(p_k, r_k, m, \beta)$\;
}

\BlankLine
Compute final residue: $r_{\text{final}} \gets \max_{y} |\widetilde{p}_{N} - p_{N}|$\;
Record trial $t = (\boldsymbol{\theta}, \text{value}=r_{\text{final}}, M=M)$ in $\mathcal{T}$\;
Update Optuna model
}

\BlankLine
\tcp{Return set of hyperparameters with residue $r \leq \epsilon$ and with the highest value of $M$}
$\mathcal{T}_{\text{valid}} = \{ t \in \mathcal{T} \, | \, t\text{.value} \leq \epsilon \}$

\BlankLine
\eIf{$\mathcal{T}_{\text{valid}} \ne \emptyset$}{
$t^* \gets \arg\max_{t \in \mathcal{T}_{\text{valid}}} t.M$\;
$\boldsymbol{\theta}^* \gets t^*.\boldsymbol{\theta}$\;
\Return{$\boldsymbol{\theta}^*$}\;
}{
\Return{\textbf{null}}\;
}
\end{algorithm}
\vspace{0.5cm}

\begin{figure}[H]
\centering
\fbox{\includegraphics[width=0.92\linewidth]{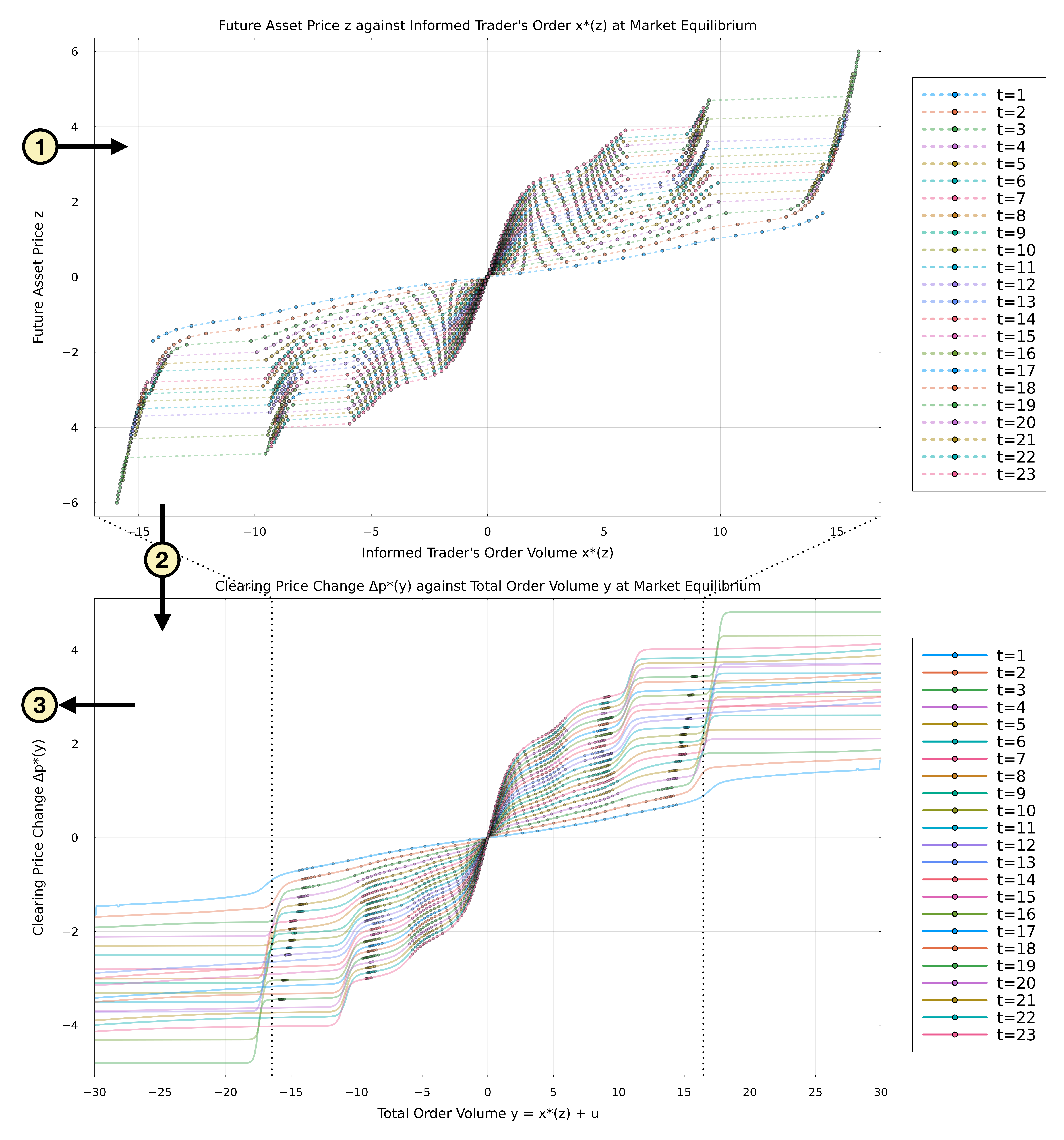}}
\caption[]{
Numerically-obtained Market Equilibrium $(x^*,p^*)$ curves for the Kyle model in which the future price of the asset is modeled by a continuous-time quantum walk. We computed the equilibrium for $t \in \{1,\cdots,23\}$. The yellow circles indicate the sequence of events in the Kyle model. Initially, the value of $t$ is fixed. 
Explanations for steps \begin{tikzpicture}
\node[draw, circle, inner sep=1pt, minimum size=1.0em] at (0,0) {1};
\end{tikzpicture}
\begin{tikzpicture}
\node[draw, circle, inner sep=1pt, minimum size=1.0em] at (0,0) {2};
\end{tikzpicture}
\begin{tikzpicture}
\node[draw, circle, inner sep=1pt, minimum size=1.0em] at (0,0) {3};
\end{tikzpicture}
are as in Figure \ref{fig:classical_equilibrium_combined}.
(Top) The plot represents the informed trader's equilibrium strategy $x^*(z)$ given a future asset value $z$. The dotted lines are only for visual aid. (Bottom) The plot represents the change in the market maker's pricing rule $p^*(y)$ for every total order volume $y$. We have defined $\Delta p^*(y)$ as $ p^*(y) = p_0 + \Delta p^*(y)$. The circles on the plot indicate the possible orders of the informed trader.
}
\label{fig:quantum_equilibrium_combined}
\end{figure}

We present the market equilibrium for the Kyle model with continuous-time quantum walk in Figure \ref{fig:quantum_equilibrium_combined}.

\begin{figure}[H]
\centering
\fbox{\includegraphics[width=0.9\linewidth]{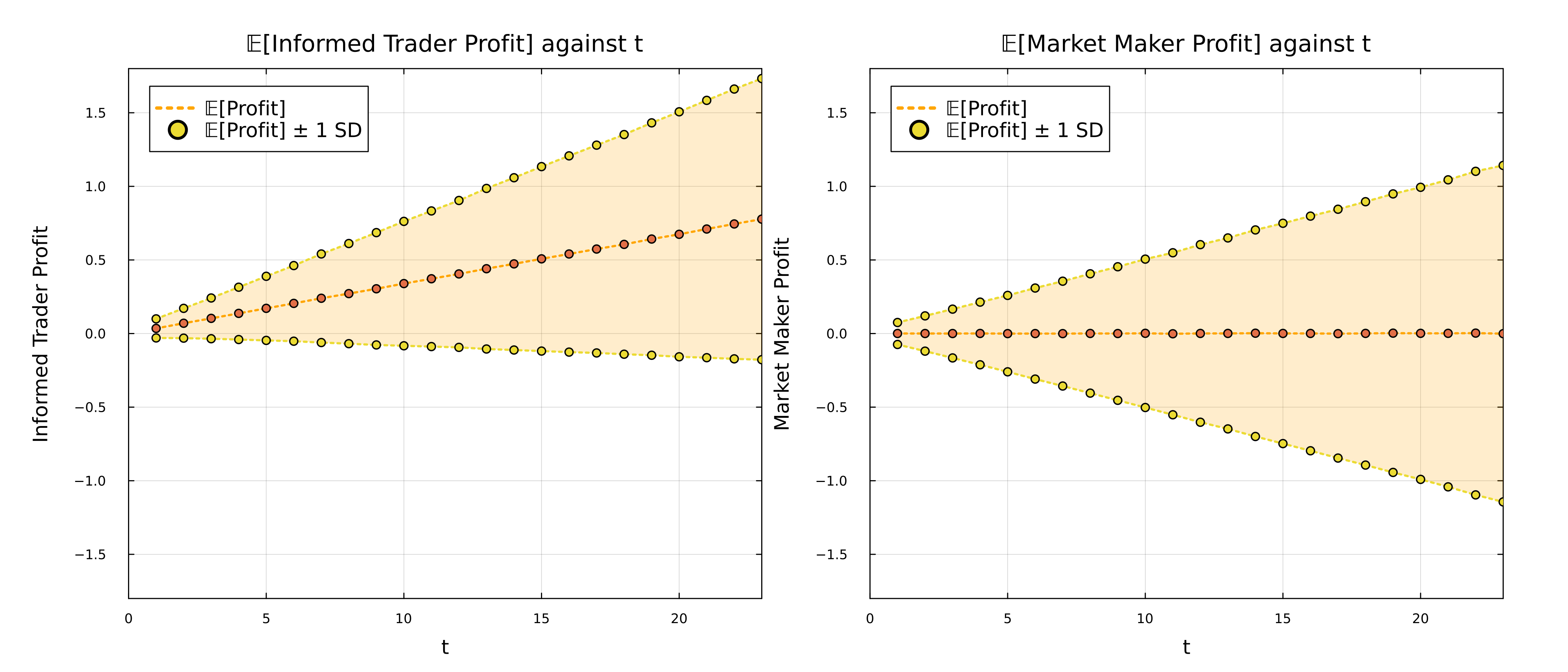}}
\caption{Average profit per round for $t \in \{1,...,23\}$ based on $200,000$ rounds of the Kyle Model for (Left) Informed Trader (Right) Market Maker}
\label{fig:quantum_equilibrium_profits}
\end{figure}

\begin{figure}[H]
\centering
\fbox{\includegraphics[width=0.9\linewidth]{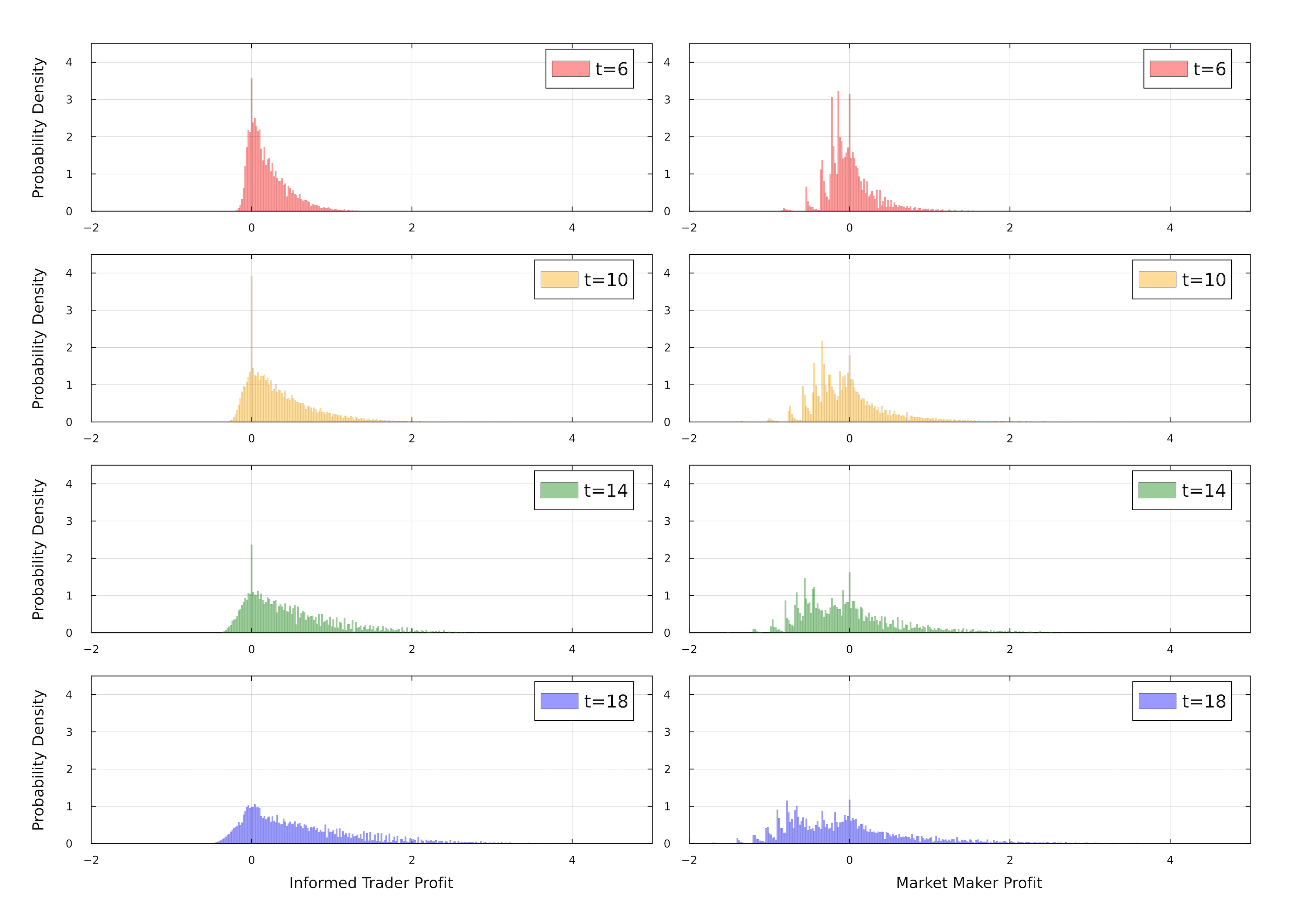}}
\caption{Probability density based on $200,000$ rounds of the Kyle Model for $t \in \{6,10,14,18\}$ for (Left Column) Informed Trader (Right) Market Maker}
\label{fig:quantum_profit_grid.png}
\end{figure}

\newpage
\subsection{Discussions}

In the continuous-time quantum walk framework where $t \in \{1,\cdots,23\}$, the probability of the future price of the asset being $z_j$ is given by $\mathbb{P}[z=z_j] = J_j^2(t) > 0 \,\,\, \forall j \in \mathbb{Z}$ \cite{bessel_transcendentality}. In our computation, we had only considered a finite set of future asset prices $\{z_j\}_{-M \,:\, 1 \,:\, M}$. This is because we empirically observed that Algorithm \ref{alg:clearing_price_equilibrium} fails to converge for large values of $M$. As such, we determined the value of $M$ using hyperparameter finetuning. From this process, we obtained values of $M$ that satisfied
\begin{equation*}
\int_{j=-M}^{M} \mathbb{P}[z=z_j \,|\, t] > 1 - 10^{-8} \quad \forall t \in \{1,...,23\}.
\end{equation*}
which we deem to be sufficient.

From the market equilibrium illustrated in Figure \ref{fig:quantum_equilibrium_combined}, we observe that the informed trader's strategy exhibits several distinct clusters with large gaps between them. These gaps correspond to jumps that we observe in the market maker's clearing price. We remark that the market equilibrium clearing price bears some resemblance to the clearing price derived from our analytical iteration as illustrated in Figure \ref{fig:analytical_clearing_price}. Finally, from the clearing price, we identified that Kyle Lambda grows linearly with $t$ and can be described by $\lambda(t) \approx 0.0517 \,t - 0.00905$ with $R^2 \approx 0.9999$. 

Despite the market equilibrium becoming increasingly non-linear with $t$, we observed that the average profits of the informed trader with respect to $t$ remain remarkably linear as illustrated in  Figure \ref{fig:quantum_equilibrium_profits}. By fitting a linear regression model, we found that the line can be described by $\E[\text{informed trader profit}] \approx 0.0407 \, t + 0.0159$ with $R^2 \approx 0.9999$.



\section{Discussion and Conclusion}

There are a few key differences between the continuous-time quantum setting and the classical walk setting. The Kyle lambda grows quadratically faster with $t$ ($\lambda(t) \propto t$) in the quantum setting than in the classical setting ($\lambda(t) \propto t^{0.491}$. Similarly, the average profit per round in the quantum setting increases linearly with $t$ compared to $t^{0.69}$ in the classical setting. Furthermore, the market equilibrium is found to be highly non-linear in the quantum setting as compared to the classical setting where the informed trader's strategy was linear.

In this paper, we have incorporated the continuous-time quantum walk into the Kyle model and identified a highly non-linear market equilibrium. Our work demonstrates how ideas from quantum mechanics could be applied to insider trading models. We believe that future work in this direction may yield new results, both theoretical and empirical, which are not seen in the classical regime.

\section{Acknowledgements}

Shawn Liew Hong Wei is supported by the National Research Foundation, Singapore, through the National Quantum Office, hosted in A*STAR, under its National Quantum Scholarship Scheme (NQSS) Funding Initiative (PhD).
This project is supported by the National Research Foundation, Singapore, through the National Quantum Office, hosted in A*STAR, under its Centre for Quantum Technologies Funding Initiative.

\printbibliography

\end{document}